\documentclass[aps,prd,twocolumn,floatfix,nofootinbib]{revtex4}
\usepackage{graphicx} \usepackage{epsf}
 
\def\gsim{ \lower .75ex \hbox{$\sim$} \llap{\raise .27ex \hbox{$>$}} }
\def\lsim{ \lower .75ex \hbox{$\sim$} \llap{\raise .27ex \hbox{$<$}} }
\def\be{\begin{equation}}  \def\ee{\end{equation}}

\def\Trh{T_{\mbox{\tiny rh}}}
\def\mpl{m_{\mbox{\tiny Pl}}}

\newcommand{\myfigure}[2]{\resizebox{#1}{!}{\includegraphics{#2}}}

\begin{document}

\title{Stochastic Gravitational Wave Production After Inflation}

\author{Richard Easther}  \author{Eugene A. Lim}
 \affiliation{Department of Physics, Yale
University, New Haven  CT 06520, USA \\ Email: {\tt
richard.easther@yale.edu} {\tt eugene.lim@yale.edu}} 

\begin{abstract} 
In many models of inflation, the period of accelerated expansion ends with preheating, a highly non-thermal phase of evolution during which the inflaton pumps energy into a specific set of momentum modes of field(s) to which it is coupled.  This necessarily induces large, transient density inhomogeneities which can source a significant spectrum of gravitational waves.  In this paper, we consider the generic properties of gravitational waves produced during preheating, perform detailed calculations of the spectrum for several specific  inflationary models, and identify problems that require further study.   In particular,  we argue that if these gravitational waves  exist  they will necessarily fall within the frequency range that is feasible for direct detection experiments -- from laboratory through to solar system scales. We extract the gravitational wave spectrum from numerical  simulations of preheating after $\lambda \phi^4$ and $m_{\phi}^2 \phi^2$ inflation, and find that they  lead to a gravitational wave amplitude of around $\Omega_{gw}h^2\sim 10^{-10}$. This is considerably higher than the amplitude of the primordial gravitational waves produced during inflation. However, the typical wavelength of these gravitational waves is considerably shorter than LIGO scales, although in extreme cases they may be visible at scales accessible to the proposed BBO mission. We survey possible experimental approaches to detecting any gravitational wave background generated during preheating.  
\end{abstract} 
 
\maketitle

\section{Introduction}
 
 In recent years, the Cosmic Microwave Background (CMB) has been our primary window into the primordial universe. Scale invariant density perturbations, sourced by quantum fluctuations during inflation and processed by the photon-baryon plasma, are by far the most satisfying explanation for the observed temperature anisotropies in the CMB. Inflation also predicts the existence of a scale invariant spectrum of primordial gravitational waves, sourced by the same quantum fluctuations  that underlie the scalar perturbations. Gravitational waves are only weakly coupled to matter fields, and move freely through the universe from the moment they are produced. They are thus the deepest probe of the early universe of which we currently know \cite{Starobinsky:1979ty,Rubakov:1982df,Fabbri:1983us,Abbott:1984fp,Allen:1987bk,Turner:1993vb,Boyle:2005se}. 

In the short term, the best hope for observing primordial gravitational waves is via their contribution to the B-mode of the CMB polarization. However, foreground weak lensing puts a fundamental limit on a B-mode signal sourced by gravitational waves \cite{Knox:2002pe,Hirata:2003ka,Seljak:2003pn}. Consequently,  attention has recently focussed upon direct detection experiments \cite{Boyle:2005se,Ungarelli:2005qb,Crowder:2005nr,Smith:2005mm} which, while enormously challenging, may ultimately be more sensitive to a stochastic  gravitational wave background than the CMB B-mode.  These experiments are sensitive to frequencies far higher than those that contribute to the CMB, since the physical sizes of detectors  necessarily range from laboratory scales through to solar system  scales.  However, as the inflationary spectrum is almost scale-free, the amplitude at short scales is not  dramatically different from that seen at CMB scales, at least for simple models of inflation.

The primordial spectrum is not alone: gravitational waves are produced whenever there are large, time-dependent inhomogeneities in the matter distribution. These more pedestrian gravitational waves are generated ``classically'', much like radiation produced by the oscillation of an electron, whereas the inflationary  tensor perturbations are sourced by quantum fluctuations in the spacetime background.  During the evolution of the universe, there are several hypothetical mechanisms which would generate large, local inhomogeneities. These include first order phase transitions \cite{Kamionkowski:1993fg}, brane-world models \cite{Sahni:2001qp,Sami:2004xk}, networks of cosmic strings \cite{Vilenkin:1981bx}, or inhomogenous neutrino diffusion \cite{Dolgov:2001nv}. Here, we investigate a further possibility:  gravitational waves produced during preheating  
\cite{Felder:2000hq,Garcia-Bellido:1997wm,Greene:1997ge,Greene:1997fu,Traschen:1990sw,Kofman:1994rk,Greene:1998nh,Giudice:1999fb,Khlebnikov:1997di,Garcia-Bellido:1998gm,Kofman:1997yn,Bassett:1998wg,Easther:1999ws,Parry:1998pn,Finelli:1998bu,Liddle:1999hq,Bassett:2005xm,Suyama:2004mz,Tsujikawa:2002nf,Finelli:2001db,Tilley:2000jh,Henriques:2003ga}, a period of non-thermal evolution following the end of inflation.     

While the detailed dynamics of resonance and preheating is a complicated, nonlinear problem, the basic picture is very simple.  For a given combination of fields, parametric resonance occurs when some subset of Fourier modes have exponentially growing solutions, driven by the oscillating inflaton  field. The resonant modes are  quickly pumped up to a large amplitude $\Phi^2 \approx \langle \chi^2 \rangle$. Resonance ends when nonlinearities render further growth kinematically expensive, leaving the  universe  far from thermal equilibrium. Thermalization occurs as the excited modes dissipate their energy over the entire spectrum via self-interaction. The rapid rise in the mode amplitude can be associated with an exponentially growing occupation number (at least for bosonic species). If one transforms into position-space, the highly pumped modes correspond to large, time dependent inhomogeneities, ensuring the matter distribution has a non-trivial quadrupole moment, sourcing the production of gravitational radiation.  

 This topic was first addressed by Khlebnikov and Tkachev \cite{Khlebnikov:1997di},  and has not been widely discussed within the experimental community. We believe that the time is ripe for revisiting this question.   Since \cite{Khlebnikov:1997di} was written, technological approaches to gravitational wave detection have advanced considerably. Moreover, there have been significant advances in the theoretical understanding of preheating in more complicated inflationary models. Finally,  numerical simulations benefit from the gains in computational power over this interval.    It will turn out that for ``typical''  cosmological parameters, gravitational radiation sourced by preheating has a peak frequency in the MHz band. Coincidentally, a new generation of detectors has been proposed which is tuned to gravitational waves in this range \cite{Ballantini:2005am}, although their strain sensitivity would need improve over current values by around $10^6$ in order to detect the spectra we compute here. This is clearly ambitious, but probably not unreasonable when compared to the 25 year interval anticipated before BBO [Big Bang Observer] will be in a  position to detect the inflationary  gravitational wave spectrum \cite{Phinney}.  

Since preheating occurs in many (although not all) inflationary models, any gravitational wave signal associated with preheating would provide a new and currently unexplored window into inflationary physics. It is worth emphasizing that the physical principles that underlie the calculations in this paper are well established and do not rely on exotic physics -- in particular the mechanism that governs the generation of the gravitational waves is simply the usual quadrupole related emission.\footnote{The possibility that preheating dynamics are directly affected by ``back-reaction'' from metric perturbations has also been discussed \cite{Finelli:1998bu,Parry:1998pn,Easther:1999ws,Bassett:1998wg,Finelli:2001db,Suyama:2004mz}  but we do not address this effect in this work.} Finally, this signal carries information about the epoch at the end of inflation, opening a new window into the early universe. With the stakes this high any possible observational opportunity must be carefully explored. 

Unlike the primordial spectrum, gravitational waves induced by preheating (or any subsequent phase transition) will not be scale-free; indeed their spectrum is indicative of the complicated processes that generate them. This is both a boon and a possible pitfall. A scale dependent spectrum necessarily contains more information than  a scale-free spectrum, so the detection and subsequent mapping of a gravitational wave background induced by preheating would yield a  rich trove of information.  However,  gravitational wave detectors are subject to physical limitations, since they all ultimately measure  deformations in (an array of) physical objects induced by passing gravitational waves.  Thus, even in principle, it is hard to imagine the direct detection of gravitational waves below atomic scales, or beyond solar system scales. To be sure, this is a large range but it is very much shorter than any relevant cosmological scale. Fortunately, by a happy numerical coincidence, any gravitational wave spectrum generated during preheating will peak at scale between a few meters and millions of kilometers -- which overlaps with the ``golden window'' open to direct detection experiments.  

In Section (\ref{sect:preheating}), we briefly review preheating and discuss previous work. Following that, in Section (\ref{sect:lengthscales}) we demonstrate that preheating gravitational waves from inflation scales ranging from TeV to the GUT scale will peak around $1$ Hz to $10^8$ Hz. In Section (\ref{sect:production}) we discuss our methodology for numerically computing the gravitational wave spectrum. In Section (\ref{sect:results}) we present our results for $\lambda \phi^4$ and $m^2 \phi^2$ inflation, confirming and extending the results of \cite{Khlebnikov:1997di}. We note that while the inflationary dynamics of the two models are very similar, their resonant behaviors diverge considerably.   Finally, we discuss the theoretical and observational implications of our results and lay out future plans to further improve the computational methodology in Section (\ref{sect:conclusions}).

  \section{Preheating} \label{sect:preheating}

A key problem facing any inflationary model is to ensure that inflation ends. This issue is highlighted in Guth's foundational paper, describing what is now known as old inflation, which is driven by a meta-stable false vacuum does not successfully terminate \cite{Guth:1980zm}.  In new or chaotic (slow-roll) inflation \cite{Linde:1981mu,Albrecht:1982wi,Linde:1983gd},  it was thought that inflation was followed by a period of {\em reheating\/}, where energy  slowly bleeds from the inflaton field as it oscillates about the minimum of its potential \cite{Albrecht:1982mp}.  Generating the correct perturbation spectrum typically requires that inflaton self-coupling is extremely weak, and this small coupling must be protected from loop corrections. Consequently, the coupling between the inflaton and other particles is necessarily tiny ($\lesssim 10^{-6}$ for $\lambda \phi^4$), ensuring that reheating proceeds slowly.

Preheating provides a vastly more efficient mechanism for extracting energy from the inflaton field, and it proceeds non-perturbatively and non-thermally via a process known as parametric resonance \cite{Traschen:1990sw,Kofman:1994rk}.  This is akin to stimulated emission in  a laser: during preheating, individual momentum modes  of fields coupled to the inflaton (or the inflaton itself, in some cases) have exponentially growing amplitudes.    Consider the following action
\begin{eqnarray}
S&=&\int dx^4 \sqrt{-g}\left[\frac{\mpl ^2 R}{16\pi }-\frac{1}{2}(\partial \phi)^2 - V(\phi)- \right. \nonumber \\ 
&&\left.\frac{1}{2}(\partial \chi)^2-\frac{1}{2}g^2\phi^2\chi^2\right]. \label{eqn:action}
\end{eqnarray}
As usual the Hubble parameter, $H$ and scale factor, $a$, are related by  $H = \dot{a}/a$ and during inflation the dynamics are described by%
\begin{eqnarray}
H^2 &=& \frac{8 \pi}{3 \mpl^2} \left[ \frac{\dot{\phi}^2}{2} + \frac{\dot{\chi}^2}{2}  + {\cal V}(\phi,\chi) \right] \, ,  \\
 \dot{H} &=& -\frac{8 \pi}{ \mpl^2} \left[ \frac{\dot{\phi}^2}{2} + \frac{\dot{\chi}^2}{2}  \right] \, , \\
\ddot{\phi} &+& 3 H \dot{\phi} + \frac{\partial {\cal V}}{\partial \phi} =0 \, ,\\
\ddot{\chi} &+& 3 H \dot{\chi} + \frac{\partial {\cal V}}{\partial \chi} =0 \, ,
\end{eqnarray}
where ${\cal V}=V(\phi)+1/2g^2\phi^2\chi^2$ is a shorthand for the potential terms in (\ref{eqn:action}). For simplicity we consider only couplings to scalar fields although fermionic preheating has also been investigated \cite{Greene:1998nh,Giudice:1999fb}.

In this action, $\phi$ is the inflaton and the inflationary dynamics are fixed by the potential $V(\phi)$, which is assumed to possess  a minimum.  At the end of inflation,  the potential energy of the field is quickly converted into kinetic energy, and the field oscillates with frequency $m_{\phi} = \sqrt{d^2V(\phi)/d\phi^2}$ evaluated at the minimum of $V(\phi)$. The solution for $\phi$ is then approximately
\begin{equation}
\phi(t)=\Phi(t)\sin m_{\phi} t.
\end{equation}
Meanwhile, the equation of motion for the $\chi$ field after expanding it in Fourier modes is  \cite{Kofman:1994rk}
\begin{equation}
\chi_k''+3H\chi_k'+(A(k)-2q\cos 2z)\chi_k=0 \label{eqn:eom}.
\end{equation}
In the limit where we can ignore the expansion of the universe this is the Mathieu equation which is simply a harmonic oscillator with a periodic forcing function. Here we have made the identification $A(k)=k^2/(m_{\phi}^2 a^2)+2q$, rescaled the time to $z=m_{\phi}t$ and used a prime  to denote differentiation with respect to $z$. The crucial \emph{resonance parameter} is
\begin{equation}
q=g^2\Phi^2/(4m_\phi^2) \, .  \label{eqn:resonanceparameter}
\end{equation}
The Mathieu equation possesses both oscillatory and exponential solutions; for each individual mode $k$ one can compute $A(k)$ and $q$ to determine whether or not it goes into resonance \cite{Abramowitz}. Roughly speaking, for broad resonance where a large number of modes are excited we need $q>1$ \cite{Kofman:1997yn}.

As first described by \cite{Traschen:1990sw,Kofman:1994rk}, some $\chi_k$ will have exponentially growing solutions for realistic parameter values. A full treatment of   \emph{parameteric resonance} in an expanding universe is complicated,  but we can make several generic statements. Firstly, preheating is very efficient  and proceeds much more rapidly than reheating, which relies on tree level couplings between the inflaton and other matter fields (which, at minimum are provided by gravitational interactions). Parametric resonance typically lasts less than a Hubble time, and in some models will complete in a few oscillations of the inflaton. This is because the resonant modes are rapidly pumped up to an amplitude $\langle \chi \rangle^2 \approx \Phi^2$, cutting off resonance as it becomes kinematically expensive. Once preheating ends, the pumped-up modes dissipate their energy via self-interaction with other modes, thermalizing the universe. 

Preheating leads to an initially non-thermal distribution of energy in the $\chi_k$ states. At high frequencies, corresponding to Fourier modes that are much shorter than the size of the post-inflationary Hubble horizon,  the effective mass of the $\chi_k$ state is much larger than the   function amplitude and resonance does not occur.  Meanwhile, at low frequencies, causality ensures that modes longer than the Hubble horizon are unlikely to be in resonance.\footnote{It is actually not impossible to have resonance at very small $k$, but it does not occur in generic models of preheating. For a summary see \cite{Tsujikawa:2002nf}.}  Consequently, we expect the spectrum to be narrow and centered around a wavelength which is dependent on the energy scale at the end of inflation. For the models we study in detail here, the gravitational wave spectrum induced by preheating peaks at scales  $1\sim 2$ orders of magnitude shorter than the energy scale at the end of inflation.  This comoving scale can be converted into a physical scale in the present universe once the post-inflationary behavior of the scale factor $a(t)$ is specified.   In Section (\ref{sect:lengthscales}) we show that the peak wavelength has a physical wavelength that scales as $H_e$
\begin{equation}
l_0 \propto \frac{1}{\sqrt{H_e}} \propto  \frac{1}{ V(\phi_e)^{1/4}} . \label{eqn:positionscaling}
\end{equation}
where the $e$ subscript denotes the value of $\phi$ and $H$ at the end of inflation. Lowering the inflation scale reduces the reheating scale, and reddens the gravitational wave spectrum.  If the longest possible modes are excited in GUT scale inflation, the signal peaks around $10^{7} \sim 10^{8}$ Hz, although the excited modes are generally slightly shorter. As we will see below, this is a very challenging frequency range for any direct detection experiment. Reducing the inflation scale pushes the signature towards more easily observable frequencies.   Given that parametric resonance naturally cuts off at both small and large scales,   the spectrum of any gravitational waves will cover a fixed range of wavelengths.  As the power is thus restricted to a relatively narrow band, the total gravitational radiation remains safely below the bound from big bang nucleosynthesis \cite{Maggiore:1999vm}.

The bottom line is that  for a short moment the universe is highly inhomogenous, providing a fertile ground for the generation of gravitational waves. Needless to say, preheating is a highly non-linear process  and analytical estimates can only take us so far. Fortunately, given an action such as (\ref{eqn:action}), this is a problem that can be solved numerically; we simply derive the equations of motion and evolve them numerically on an expanding lattice. This mimics the growth of the universe, but ignores the back-reaction of metric perturbations on the field evolution -- an assumption that is self-consistent, as while $\delta \rho/\rho$ can be large, the metric perturbations are typically small.  We use a modified version of the publicly available package {\sc LatticeEasy} \cite{Felder:2000hq} for the numerical computations.

To our knowledge, the generation of gravitational waves by preheating has been thoroughly examined only once before, by  Khlebnikov and Tkachev  \cite{Khlebnikov:1997di}.   Their work is the starting point for this paper: we elaborate and expand upon their treatment, considering a broader range of models, and taking recent developments in  detector technology into account. In particular, preheating can occur at a very broad range of scales,  for example via hybrid inflation \cite{Linde:1991km,Garcia-Bellido:1997wm,Garcia-Bellido:1998gm}. If the scale gets low enough, the peak wavelength can be close to the scales probed by next generation observatories such as BBO \cite{Phinney}. In Section (\ref{sect:results}), we reproduce numerical results of \cite{Khlebnikov:1997di} for the $\lambda \phi^4$ model, and present new results for the $m^2 \phi^2$ model.  While the inflationary behavior of these two models is similar, their resonance structure is very different, providing a useful crosscheck on the generality of our analytic estimates.  In future work we will extend these numerical calculations  to hybrid inflation and fermionic preheating.

\section{Peak wavelength and amplitude} \label{sect:lengthscales}

We begin our detailed analysis with a general discussion of the different parameters that determine the amplitude and wavelength of any gravitational waves generated during preheating.  Consider the ``usual'' gravitational wave power spectrum   generated by quantum fluctuations of the background,
\begin{equation}
\Omega_{gw,inf}(k)h^2 = \Omega_{r}h^2\frac{32}{9}\left(\frac{V_e}{\mpl^4}\right) \left(\frac{g_0}{g_*}\right)^{1/3}, \label{eqn:quantumgrav}
\end{equation}
where $V^{1/4}_{inf}$ is the energy scale of inflation and $\Omega_r h^2\approx 4 \times 10^{-5}$ is the total density of radiation today.  The effective number of degrees of freedom in the radiation at matter-radiation equality and today are given by $g_*$ and $g_0$ respectively. This form of the power spectrum is slightly non-standard as tensor modes are usually expressed via $P_h=8\pi H^2/\mpl^2$. There are two salient features to this spectrum. The first is that it is (almost) scale-invariant, since each mode is frozen out at an approximately constant energy scale, namely the inflation scale. The second is that the power is minute; for the most optimistic scenario where inflation occurs around the GUT scale, $\Omega_{gw,inf}(k) h^2 < 10^{-14}$. The current upper limit on the scale of inflation from WMAP observations for single-field inflation models is $V^{1/4}<3.3 \times 10^{16}$ GeV, corresponding to $\Omega_{gw,inf}h^2<2\times 10^{-15}$ \cite{Peiris:2003ff,Smith:2005mm}

The gravitational waves which we are considering in this paper are not directly sourced by quantum fluctuations; instead they are generated by the classical motion of particles during preheating. As is well-known,  accelerated motion generates a quadrupole moment, leading to the generation of gravitational radiation.     During preheating at the end of inflation, large inhomogeneities in the matter fields are generated by the selective pumping of modes in parametric resonance. These large inhomogeneities, as first shown in \cite{Khlebnikov:1997di} for the $\lambda \phi^4$ model, are sufficiently large to produce gravitational waves with amplitudes many orders of magnitude larger than those produced by the quantum fluctuations.  An analytical estimate \cite{Khlebnikov:1997di}, for an inflaton with an effective oscillation frequency $\bar{m}$, coupled to a massless scalar field $\chi$ with a $g^2\phi^2\chi^2$ term, yields the peak amplitude at the resonance mode $k \sim H_r$
\begin{equation}
\Omega_{gw}(k\sim H_e)h^2\approx \Omega_{r}h^2\frac{\bar{m}^2}{g^2 \mpl^2}\left(\frac{g_0}{g_*}\right)^{1/3}. \label{eqn:simpleamplitude}
\end{equation}
In other words, the amplitude probes the oscillation scale,   in contrast to the primordial spectrum which probes the inflation scale. If we plug in the usual field values for chaotic inflation $\bar{m}=m_{\phi}$ at the end of inflation $\phi\approx \mpl$ such that $V_e=m_{\phi}^2\phi^2/2 = m_{\phi}^2\mpl^2 /2 $, we see from equation (\ref{eqn:quantumgrav}) that the amplitude of the gravitational waves generated by preheating is $1/g^2$ larger than the inflationary spectrum. This is a significant boost, as we expect $g^2 \lesssim 10^{-6}$.  Note that  in  models such as hybrid inflation \footnote{We note that in hybrid inflation, preheating amplification of the perturbations is achieved through a combination of parametric resonance and tachyionic instabilities. We thank Gary Felder for pointing this out to us.} one can decouple the oscillation frequency $\bar{m}$ from the inflaton mass $m$, and this simple relationship has to be revisited \cite{Garcia-Bellido:1997wm,Garcia-Bellido:1998gm}. 

However, only a finite range of modes excited during preheating. If the power was generated at scales corresponding to, say, atomic distances today, then our hope of detecting any  gravitational waves induced by preheating  would be dashed. On causal grounds,  we expect that resonant modes have a wavelength roughly equal to or less than the Hubble  length at the end of inflation, $1/H_e$:
\begin{equation}
H_e \sim \frac{ \sqrt{V_e}}{\mpl} \, ,
\end{equation} 
where $V_e$ is the inflationary potential. After inflation, the universe reheats to a temperature $\Trh$. During the subsequent radiation dominated phase, the Hubble parameter scales as $H = H_*(a_*/a_e)^2$ until matter-radiation equality at $T_*$.   Meanwhile, the scale factor evolves as $a_* = a_0 (g_0/g_*)^{1/2} (T_0/T_*)$ from matter-radiation equality until today when $a_0\equiv 1$. Thus for a physical length $l$ \cite{Khlebnikov:1997di} and \emph{physical} wavevector $k$ we have
\begin{eqnarray}
l&=&\frac{1}{k} \frac{g_*^{1/2}}{g_0^{1/3}}\left(\frac{8\pi^3}{90}\right)^{-1/4}\frac{\sqrt{H_e M_p}}{T_0} \nonumber \\
& \approx& 0.5\frac{\sqrt{M_p H_e}}{k} ~\mathrm{cm} 
\end{eqnarray}  
or
\begin{equation}
f=6 \times 10^{10} \frac{k}{\sqrt{M_p H_e}} ~\mathrm{Hz} \label{eqn:frequency}
\end{equation}
where we have used $g_0/g_* = 1/100$ in the second line. Plugging in the lowest excitable frequency $k = H_e$, where we more or less expect peak gravitational waves production to occur, we obtain the scaling relation (\ref{eqn:positionscaling}), as claimed earlier.

The inverse scaling is particularly important: it means that the pertinent wavelengths are longer for smaller inflationary energy scales. If we assume instantaneous reheating after inflation for GUT scale inflation $H_e \approx 10^{13}$ GeV, $l\approx 1-10$ meters, and $f\approx 10^{7}$ to $10^{8}$ Hz.  Lowering the inflationary scale reduces power in the primordial gravitational wave spectrum  making it  harder to detect, as quantified by equation (\ref{eqn:quantumgrav}). However, this also reddens the peak power of any preheating generated gravitational waves, making them easier for us to observe. This follows because the strain sensitivity $\tilde{h}_f$ of a detector scales as $\Omega_{gw}/f^3$ \cite{Maggiore:1999vm}, i.e. for the same value of $\Omega_{gw}$ we have to build a more sensitive detector if the frequencies are higher. In addition, if inflation occurs at a lower scale, then the gravitational wave energy density will be diluted less by expansion following preheating, again increasing our chance of observing them.  On the other hand, if the gravitational waves are generated at a lower scale the off-diagonal terms of $T_{\mu\nu}$ are smaller for fixed $\delta \rho / \rho$, and will be less efficient sources of gravitational radiation.  On the basis of the limited calculations performed in this paper, we see that the last two effects roughly cancel and $\Omega_{gw}$ does not depend strongly on $H_e$. However, further work will be needed before we can safely say that this is true of all models which undergo preheating. 
 
In this naive analysis,  a few subtle points have been glossed over. The peak resonance modes are usually not exactly at the Hubble scale; instead they are frequently $1\sim 2$ orders of magnitude smaller \cite{Greene:1997fu,Kofman:1997yn,Greene:1997ge}.  This has the effect of pushing the observable modes to a bluer band.  On the other hand, preheating does not always start immediately after inflation ends; peak particle production occurs when the amplitude of the field perturbations $\delta \phi/\phi$ grows to order unity, which need not happen quickly. In the models looked at here, the Hubble parameter during peak gravitational wave production is about $1\sim 2$ orders of magnitude smaller than $H_e$,  shifting the observable modes to a redder band. 

It is also worth noting that the gravitational waves are generated causally within the Hubble volume, and thus the phases of the individual modes are uncorrelated -- unlike the primordial spectrum. This is a generic feature of all causally generated perturbation spectra, and is a powerful discriminant \cite{Dodelson:2003ip}. Unfortunately, direct detection experiments cannot dinstinguish the coherence (or lack of) of the gravitational waves as their signal is an integral over some time interval greater than the frequency scale. To do this, one  must find a processed \emph{imprint} on a fixed time-slice. 
 
 While there is an upper bound on the inflationary energy scale from the contribution of tensor modes to the CMB, the lower bound is very weak.  At minimum,  the post-inflationary universe must be hot enough to permit baryogenesis and nucleosynthesis. We conservatively assume that the former occurs via electroweak scale processes, so we can easily have  $V_e^{1/4}$ as low as the TeV scale. Nuclear reactions necessarily take place at MeV scales and ensuring successful nucleosynthesis provides an absolute lower limit on the reheating temperature.    This corresponds to gravitational wave peak wavelength scales ranging from laboratory scales through to solar system scales.     

\section{Gravitational Wave Production}  \label{sect:production}

Equation (\ref{eqn:simpleamplitude}) suggests that the preheating induced gravitational wave spectrum is larger than its primordial  counterpart. To obtain an actual power spectrum, the highly nonlinear physics of preheating forces us to turn to numerical methods. We use {\sc LatticeEasy\/} \cite{Felder:2000hq} to simulate the evolution of the early universe, solving the equations of motion for a set of interacting scalar fields in a flat Friedmann-Robertson-Walker (FRW) Universe. The fields become highly inhomogeneous, which is important for the generation of gravitational waves. This does not immediately make the {\em metric\/}  perturbations large \cite{Ishibashi:2005sj}. Consequently, we can solve the nonlinear field evolution numerically while assuming a rigid spacetime background, and then extract the spectrum of gravitational radiation produced during preheating.

We extended {\sc LatticeEasy\/} to compute the gravitational wave spectrum generated during preheating. We follow the approach of \cite{Khlebnikov:1997di} (see also \cite{Kamionkowski:1993fg}), reproducing their results for the quartic $\lambda \phi^4 /4 + g^2\phi^2\chi^2/2$ model. In addition, we compute the gravitational wave spectrum for the quadratic inflation model $m^2\phi^2/2 +g^2\phi^2\chi^2/2$.  We leave the simulation of other models such as the negative coupling $-g^2\phi^2\chi^2/2$ \cite{Greene:1997ge} or hybrid inflation models to future work.  

We now sketch the approach we use to compute the spectrum.  We begin by considering the energy radiated in gravity waves in a frequency interval $d\omega$ and a solid angle $d\Omega$, given by
\begin{equation}
\frac{dE}{d\Omega}=2G\Lambda_{ij,lm}\omega^2 T^{ij*}(\vec{\mathbf{k}},\omega)T^{lm}(\vec{\mathbf{k}},\omega) d\omega \label{eqgrav}
\end{equation}
where $T^{ij}$ is the stress tensor describing the source matter fields. Here $i$,$j$ run over the spatial indices and the projection tensor is given by \cite{WeinbergBook}
\begin{eqnarray}
\Lambda_{ij,lm}(\hat{k})=\delta_{ij}\delta_{lm}-2\hat{k}_j\hat{k}_m\delta_{il}+\frac{1}{2}\hat{k}_i\hat{k}_j\hat{k}_l\hat{k}_m \nonumber \\
-\frac{1}{2}\delta_{ij}\delta_{lm}+\frac{1}{2}\delta_{ij}\hat{k}_l\hat{k}_m+\frac{1}{2}\delta_{jl}\hat{k}_i\hat{k}_m
\end{eqnarray}
with unit vector $\hat{k}\equiv \vec{\mathbf{k}}/\omega$. Strictly speaking, this formula is only valid for linearized gravity in Minkowski space. A more accurate calculation will involve solving the equations of motion for linearized gravity on a curved background. 

To see why  the use of this formula is justified, consider the gravitational wave energy emitted by a three dimensional box of \emph{conformally flat} spacetime with physical size $l\times l\times l$. The energy density in this box is 
\begin{equation}
d\rho(\omega)=8\pi G  l^{-3} \Lambda_{ij,lm}\omega^2 T^{ij*}(|\mathbf{\vec{k}}|,\omega)T^{lm}(|\mathbf{\vec{k}}|,\omega) d\omega  \label{eqn:gravdensity}
\end{equation}
where we have assumed that the spectrum is isotropic.\footnote{Isotropy allows us to choose any direction for the projection vector $\Lambda_{ij,lm}$: we chose for simplicity $\hat{k}=(1,0,0)$. We checked that this assumption is robust by showing that the final simulation results are not sensitive to different choices of direction.} From causal arguments alone, only modes of wavelengths equal to or shorter than $1/H$ will be generated, imposing a natural cut-off at long scales.  Thus, provided we choose $l\geq 1/H$ we will effectively capture the essential physics.  Depending on how efficient preheating is in a particular model, the entire phase can  last for several Hubble times. However, the gravitational waves are produced near the end of preheating, as the inhomogeneities in the fields become large. We therefore expect the gravitational wave power to be generated in a short burst, and numerical simulations confirm this suspicion. Thus it is a good approximation to assume that the gravitational wave source is localized in the box \cite{Kosowsky:1991ua}.
 
In our simulations, we begin our computations at the end of inflation, near the beginning of the parameteric resonance phase. We end our simulations when the fields are stabilized and parameteric resonance ends.   We subdivide the the spacetime into discrete 4-D boxes of spatial sizes $L^3$ and time interval $\tau=L$, where $L$ and $\tau$ are the conformal length and time respectively. Our physical box size  thus scales roughly as $a^3$. The choice of $\tau=L$ is purely operational, allowing us to fix our Fourier variables to be the conformal frequency and conformal wavevector such that $|\mathbf{\vec{k}}_{conf}|=\omega_{conf}$. One can in principal decompose the conformal time differently, but that would unnecessarily complicate matters. We fix $L$ so that during the period of gravitational wave production $aL \geq 1/H$ and the box is larger than the effective Hubble horizon.

We assume that each ``box'', labeled $\alpha$, is a localized source, and compute the total gravitational wave density produced for each box $\rho_{gw}^{(\alpha)}$ using equation (\ref{eqn:gravdensity}). We then sum them up, diluting them appropriately as follows
\begin{equation}
\frac{d\rho_{gw}(a_{e})}{d\ln \omega}=\sum_{\alpha} \frac{d\rho_{gw}^{(\alpha)}(a_{\alpha})}{d\ln \omega}\left(\frac{a_{e}}{a_{\alpha}}\right)^{4} \label{eqn:finalpower}
\end{equation}
where $a_{\alpha}$ is the scale factor taken at the middle of the box in conformal time and $a_{e}$ is the scale factor at the end of inflation. Meanwhile, for each box $\alpha$
\begin{equation}
\frac{d\rho_{gw}(a_{\alpha})}{d\ln \omega}=8\pi G \omega^3 l_{a_{\alpha}}^{-3}\Lambda_{ij,lm}T^{ij*}T^{lm} 
\end{equation}
where $l_{a_{\alpha}}^3$ is the physical size of the box at time $a_{\alpha}$ and $\omega$ is the physical frequency.

Finally, putting everything together, the total density of gravitational waves today is given by
\begin{equation}
\Omega_{gw}h^2=\Omega_r h^2 \frac{d\rho_{gw}(a_{e})}{d\ln \omega}\left(\frac{g_0}{g_*}\right)^{1/3} \label{eqn:finalpowertoday}
\end{equation}

We should mention that in equation (\ref{eqn:finalpower}) and hence equation (\ref{eqn:finalpowertoday}), we have implicitly assumed that the universe is radiation dominated at the end of preheating, which is not true for certain chaotic models.

\section{Numerical Results} \label{sect:results}

In this section, we give numerical results for the gravitational wave spectrum produced during resonance in two different models: $\lambda \phi^4$ and $m^2\phi^2$. While the inflationary dynamics of these two systems are very similar, there is considerable divergence in the resonance  structure between the models, making this a useful generalization of \cite{Khlebnikov:1997di}.

\subsection{Quartic Inflation ($\lambda \phi^4$)}

\begin{figure}[ptb]
\myfigure{3.25in}{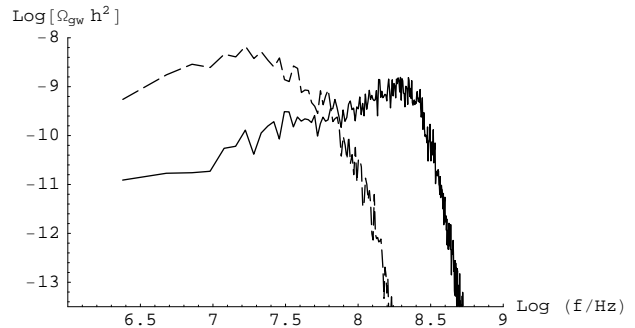}
\caption{The gravitational spectrum for the $\lambda \phi^4$ model with $\lambda=10^{-14}$ and $g^2/\lambda=1.2$ (dash line) and $120$ (full line) respectively . As expected, it is peaked around $10^7\sim 10^8$ Hz and spans about 2 decades. The horizon size at the time of preheating imposes the low frequency cut-off, while the high frequency cut-off is due to the fact that high momentum $\chi$ particles are energetically too expensive to be created. Notice that the power is roughly inversely proportional to $g^2$. }
\label{fig:quarticstdresult}
\end{figure}

To test our code, we reproduce Khlebnikov and Tkachev's results    \cite{Khlebnikov:1997di}  for  $\lambda \phi^4$ with a $\phi^2\chi^2$ term
\begin{equation}
{\cal V}(\phi,\chi)=\frac{\lambda}{4}\phi^4 + \frac{1}{2}g^2\phi^2 \chi^2.
\end{equation}
From the perspective of preheating, this model is atypical \cite{Greene:1997fu}  as it possess only a weak resonance band.   Even so, we still see significant production of gravity waves.

Following \cite{Khlebnikov:1997di}, we set $\lambda=10^{-14}$ and $g^2/\lambda=120$, corresponding to a resonance parameter $q\approx 120$ from equation (\ref{eqn:resonanceparameter}). In this model, inflation ends around the GUT scale, where $\phi_0\approx \mpl$, or $H_{end}\approx 10^{12}$ GeV. We begin our simulation on a $256^3$ size lattice from that time and run it until preheating ends around $H\approx 10^{7}$ GeV. Parameteric resonance peaks around $H_{peak}\approx 10^{8}$ GeV, and the size of the box is chosen to ensure that its physical size at this time $l \approx 1/H_{peak}$. With a $\lambda \phi^4$ potential, the background spacetime scales like a radiation dominated universe during parametric resonance.

\begin{figure}[ptb]
\myfigure{3.25in}{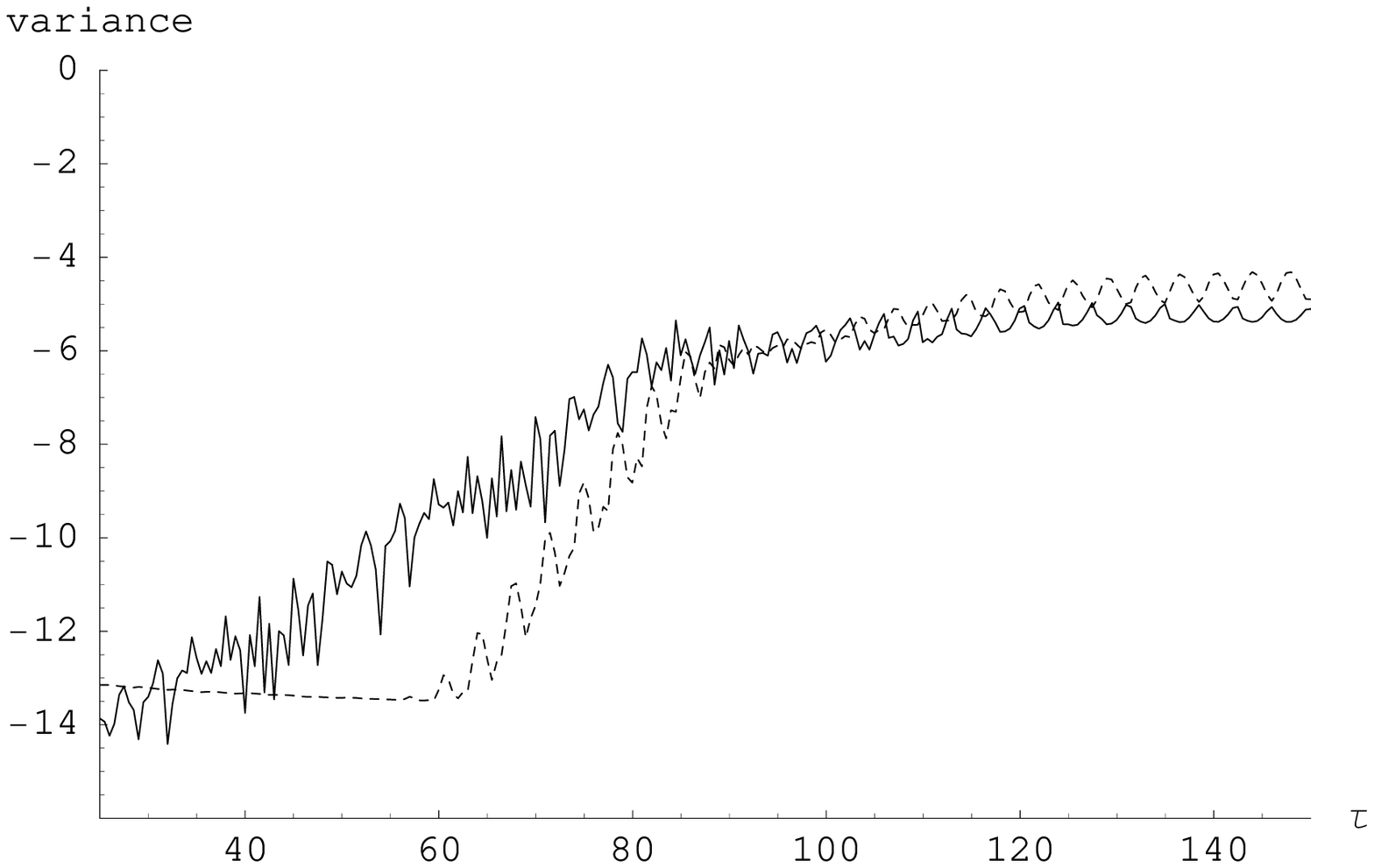}
\caption{The variance of the $\phi$ (dash) and $\chi$ (full) fields for a $\lambda \phi^4$ inflation model with  $g^2/\lambda=120$. The time coordinate is the conformal time $\tau$ in units of $2.5\times 10^{-12} \textrm{GeV}^{-1}$ while the variances are in $\mpl^2$ units. The associated hubble parameter during the rapid rise while in major resonance phase is $H\approx 10^{11}$ GeV, thus this phase lasts less than a Hubble time.} 
\label{fig:lambdavar}
\end{figure}

Using (\ref{eqn:frequency}), the present  frequency associated with the Hubble parameter during preheating is $10^{6}$Hz. From figure (\ref{fig:quarticstdresult}), we see that the peak frequency is actually $10^{7}\sim 10^{8}$ Hz, suggesting that the peak resonance modes are about two orders of magnitude smaller than the Hubble wavelength. The amplitude of the gravitational waves peaks at around $\Omega_{gw}\approx 10^{-9}$, consistent with equation (\ref{eqn:simpleamplitude}).

From the plot, we see that even at the lower end of the relevant frequency range which is easier to detect, $\Omega_{gw} \sim10^{-11}$. This is 3 orders of magnitude larger than the primordial spectrum. Beyond that at lower frequencies, we expect the spectrum to undergo a steep $k^3$ decline. This $k^3$ superhorizon  tail is a common property for a spectrum which is causally generated inside the Hubble horizon \cite{Liddle:1999hq}.  In cases where the inflationary scale and thus the intrinsic stochastic background is very low (so it  does not mask the signal)  this $k^3$ tail might be easier to detect than the peak wavelengths, given the physical limitations on realistic detectors.

\subsection{Quadratic Inflation ($m_{\phi}^2\phi^2$)}

We now turn to the $m_{\phi}^2\phi^2$ model \cite{Linde:1983gd} with the same interaction term as before
\begin{equation}
{\cal V}(\phi,\chi)=\frac{1}{2}m_{\phi}^2\phi^2 + \frac{1}{2}g^2\phi^2 \chi^2.
\end{equation}

The amplitude of the CMB temperature anisotropy requires $m_{\phi}\approx 10^{13}$ GeV. At the end of inflation $\phi \approx \mpl$. Choosing $g^2=2.5 \times 10^{-7}$, gives a resonance parameter of $q\approx 2.5\times 10^{5}$, via (\ref{eqn:resonanceparameter}). Again we begin our simulation after inflation ends at $H_{end}\approx 10^{13}$ GeV through the peak preheating phase at $H_{peak}\approx 10^{11}$ GeV, until the end of preheating. Since $m_{\phi}$ is non-zero, the universe evolves  as if it was matter dominated\footnote{The equation of state $w=\langle p \rangle / \langle \rho \rangle$ fluctuates rapidly between $1$ and $-1$ around a center value of $0$.} with the scale factor growing 30-fold. At the end of preheating, we assume that the universe reheats normally and enters a radiation dominated phase. This is in principle not a valid assumption, as some numerical results have shown that it is difficult for quadratic inflation to reheat to radiation domination without a trilinear coupling \cite{Podolsky:2005bw}. In this paper, we are using it as a toy model to illustrate preheating for a inflation model with a different mass scale.

\begin{figure}[ptb] 
\myfigure{3.25in}{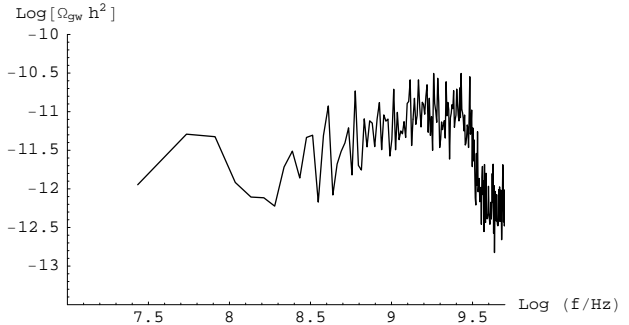} 
\caption{The gravitational spectrum for a $m_{\phi}^2\phi^2$ inflation model, with parameters $g^2=2.5 \times 10^{-7}$ and $m_{\phi}=10^{-6}\mpl \approx 10^{13}$ GeV. The resonance parameter here $q=2.5 \times 10^{5}$. The slight rise at high frequencies after the sharp drop is a numerical artifact. }
\label{fig:chaoticstdresults}
\end{figure}

We simulated this model on a $256^3$ lattice, with the results as shown in figure (\ref{fig:chaoticstdresults}).

\begin{figure}[ptb]
\myfigure{3.25in}{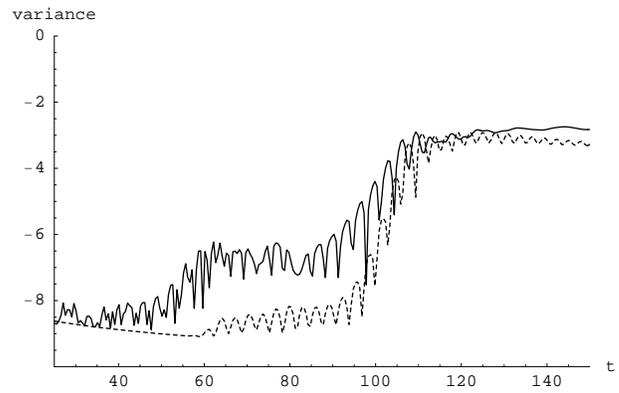}
\caption{The variance of the $\phi$ (dash) and $\chi$ (full) fields for $m_{\phi}^2\phi^2$ inflation model with $q=2.5\times 10^{5}$. The time coordinate is the cosmic time $t$ in units of $m_{\phi}^{-1}=8.33\times 10^{-14} \textrm{GeV}^{-1}$ while the variances are in $\mpl^2$ units. The associated hubble parameter during the rapid rise while in major resonance phase is $H\approx 10^{12}$ GeV, thus this phase lasts much less than a Hubble time.} 
\label{fig:chaoticvar}
\end{figure}

\begin{figure}[ptb] 
\myfigure{3.25in}{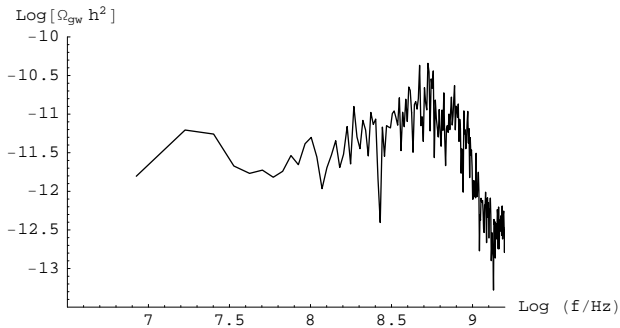} 
\caption{The gravitational spectrum for a $m_{\phi}^2\phi^2$ inflation model, with parameters $g^2=2.5 \times 10^{-7}$ and $m_{\phi}\approx10^{-7}\mpl =10^{12}$ GeV. The resonance parameter here $q=2.5 \times 10^{5}$. Although this lower mass model is ruled out by the CMB, it serves as a useful prototype to illustrate that the peak will be reddened for a lower scale inflation.}
\label{fig:chaoticstdresults2}
\end{figure}

Using (\ref{eqn:simpleamplitude}), we expect $\Omega_{gw}h^2\approx 10^{-10}$ at peak which matches the result of our detailed calculation. Although inflation ends at $H_{end}\approx 10^{13}$ GeV, peak resonance  occurs at $H_{peak}\approx 10^{11}$ GeV, which  sets the comoving size of our lattice. Figure (\ref{fig:chaoticstdresults}) shows that the peak frequency is actually $10^{8}\sim 10^{9}$ Hz, a couple of orders of magnitude smaller than the Hubble parameter during reheating.

Finally, we present the results for a model with a lower mass, $m_{\phi}=10^{12}$ GeV in figure (\ref{fig:chaoticstdresults2}). This model is ruled out by CMB data, but it demonstrates the way in which the gravitational wave spectrum generated by preheating depends on the inflationary scale. In this model  $H\approx 10^{12}$ GeV, and as expected from equation (\ref{eqn:positionscaling}), the peak location is reddened by a factor of $\sqrt{10}$. The observable power remains comparable to the previous model. The emitted power is reduced, but since the overall expansion of the universe is reduced by the lower reheating temperature, the values of $\Omega_{gw}h^2$ today is roughly fixed.  It is tempting to conjecture that the cancellation between these two effects will be seen in other preheating induced gravitational wave spectra, and that the $\Omega_{gw} h^2 \sim  10^{-10}$ seen here will prove to be a generic value \cite{Garcia-Bellido:1997wm,inpreparation}.

\section{Summary and Future Prospects} \label{sect:conclusions}

\begin{figure*}
\myfigure{7in}{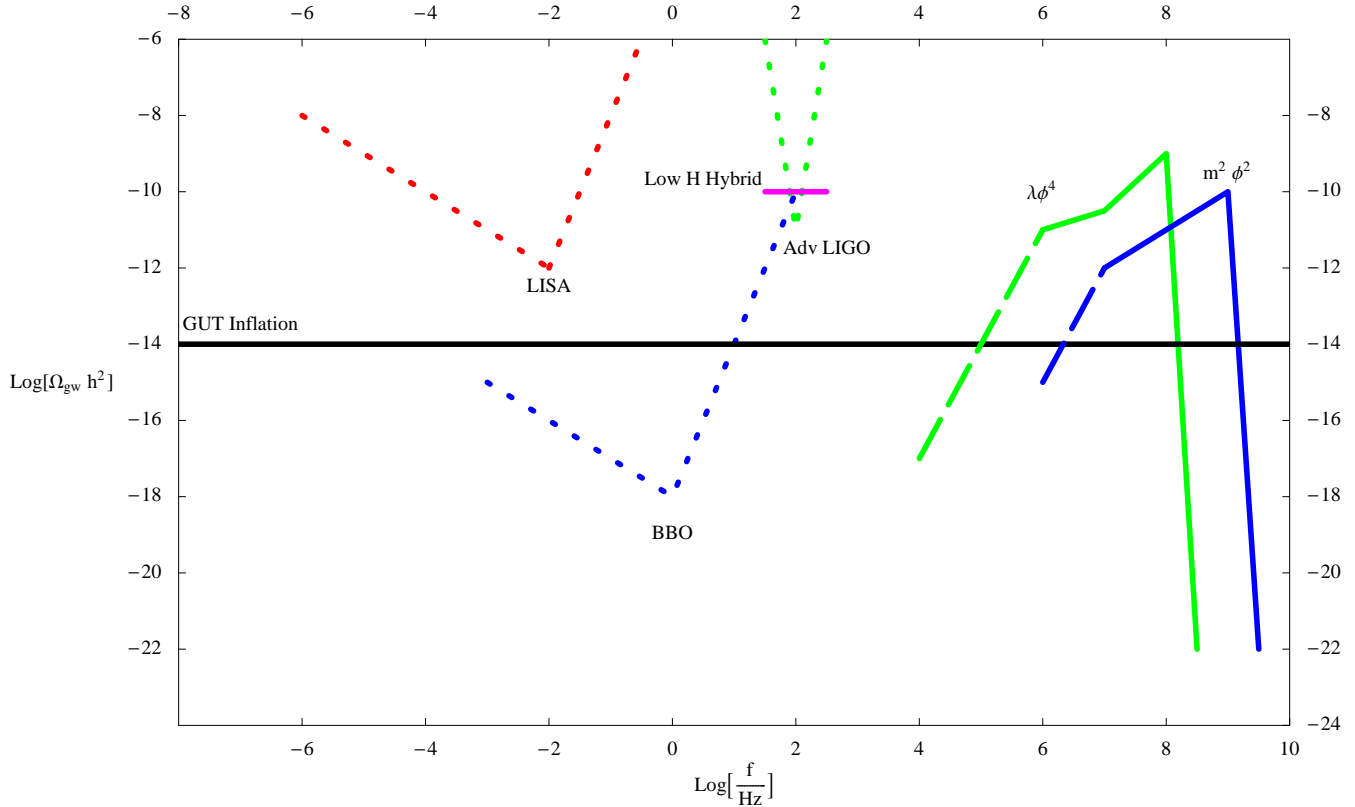}%
\caption{Projected sensitivities of gravitational wave detectors (dotted lines) to a stochastic gravitational wave spectrum (full lines) in $\log(\Omega_{gw} h^2)$ and $\log(f/\mathrm{Hz})$. The limits are for \emph{correlated} observatories, except for LISA. The primordial inflation spectrum is plotted for a scale-invariant GUT scale inflation. The gravitational wave spectra generated by the quartic and quadratic type inflation models for a GUT scale inflation are schematically shown here; see Section (\ref{sect:results}) for more detailed plots. In addition, we have extrapolated a $f^3$ tail to the red of the simulated results, which is a conservative estimate on the degree of the Hubble scale cut-off to the gravitational wave spectrum \cite{Liddle:1999hq}. We have also added the expected peak point of a TEV scale hybrid inflation model with high oscillation scale, calculated using the naive formula (\ref{eqn:simpleamplitude}) \cite{Garcia-Bellido:1998gm}.\label{fig:sensitivityplot}}
\end{figure*}

We have carefully investigated the production of gravitational waves during preheating,   reproducing the work of Khlebnikov and Tkachev \cite{Khlebnikov:1997di} for the $\lambda \phi^4$ inflation model and extending it to the $m_{\phi}^2\phi^2$ case.  For both models we show numerically that  preheating is  a sizable source of gravitational waves with frequencies of around $10^{6}\sim 10^{8}$ Hz, and peak power of $\Omega_{gw}h^2\approx 10^{-9}  \sim 10^{-11}$.  We present simple scaling arguments to predict the overall properties of the spectrum for a broader class of inflationary models. We see that the spectrum of gravitational waves induced by preheating peaks at a scale proportional to $1/\sqrt{H}$, where $H$ is the Hubble parameter during preheating, and generally somewhat smaller than the scale of inflation. Thus, lowering the inflationary scale reddens the spectrum and makes it easier to observe.  This is in contrast to the primordial inflationary spectrum, which is roughly scale invariant and becomes harder to observe as inflationary scale is lowered.

We now ask what we can learn about inflation if we detect a spectrum of gravitational waves generated during preheating. Its two most basic features, the peak frequency and the amplitude, represent the reheat and the oscillation scales respectively.  As the reheat scale is lower than the inflation scale, its detection would impose a lower bound on the inflationary scale. Its usefulness as a probe of inflation is amplified if we have a separate probe of the scale of inflation, say from the  CMB B-mode observations.  

The oscillation scale is harder to interpret, as it is often highly model dependent. For single scalar field inflation, such as the models considered here,  knowledge of the reheat scale would constrain the coupling constant $g^2$. More optimistically, the \emph{structure} of the spectrum encodes information about resonance and preheating, so if we can predict the structure accurately we potentially probe the detailed mechanics of preheating. Such an endeavour will require more careful computations and simulations than we present in this paper.

Further progress on this problem can be made in two ways \cite{inpreparation}. The first is to further refine the code to accommadate a larger class of models, particularly hybrid inflation models which have an essentially arbitrary inflationary scale. Alternatively, as alluded to in Section (\ref{sect:production}), a more sophisticated theoretical calculation would be to directly solve the evolution equations for the off-diagonal parts of the perturbed Einstein tensor, which are sourced by $T_{ij}$.  This would avoid any ambiguity concerning the use of a formula that is only strictly applicable in flat space, and it would avoid the need to run the code for a finite number of ``boxes'', since we would only need to take a Fourier transform at the end of the computation.   

Investigating these gravitational waves is timely, since there is currently considerable interest in the direct detection of gravitational waves.  At the moment, several proposals based on different technologies are being actively pursued. At the solar-system scale, the space-based interferometer LISA \cite{LISA} will probe frequencies from $10^{-2}$ Hz, which is probably too small for the gravitational waves we are considering. The proposed BBO \cite{Phinney} and also the Deci-hertz Interferometer Gravitational Wave observatory (DECIGO) \cite{Seto:2001qf} missions are sensitive to frequencies on the order of $1$ Hz, and would probe gravitational waves arising from preheating after TeV scale inflation. 

The array of terrestrial interferometers also probes frequencies corresponding to preheating following low-scale inflation. These experiments include GEO600 \cite{GEO600},  LIGO  \cite{LIGO}, TAMA \cite{TAMA} and VIRGO \cite{VIRGO}. These are sensitive to scales between $100\sim 1000$ Hz and may be able to probe a stochastic background  in the interesting range $\Omega_{gw}h^2 \sim 10^{-10}$, if they are correlated. At even higher frequencies in the KHz range, we have a slew of resonant bars detectors   \cite{ALLEGRO,AURIGA,EXPLORER,Blair:1997as}. Once correlated, these resonant bars have a potential to reach $\Omega_{gw}h^2\approx 10^{-5}$ \cite{Maggiore:1999vm},  which puts the signals we see here out of their reach. However, theoretical studies suggest that correlating hollow spherical detectors may eventually allow us to reach $\Omega_{gw}h^2\approx 10^{-9}$ \cite{Coccia:1997gy}.

At even higher frequencies from $10^{3} \sim 10^{5}$ Hz, there has been a proposal to build a superconducting resonant cavity detector called the Microwave Apparatus for Gravitational Wave Observation (MAGO) \cite{Pegoraro:1977uv,Reece:1984gv,Ballantini:2005am}. Although the strain sensitivity $\tilde{h}_f \approx 10^{-21} \textrm{Hz}^{-1/2}$ for the prototype is expected to be comparable to resonant bar detectors at $4\times 10^{3}$ Hz, the large $f^3$ suppression from the relation $\Omega_{gw} h^2 \propto \tilde {h}_f^2 f^3$ means that at these frequencies we can only reach $\Omega_{gw} h^2\approx {\cal{O}}(1)$. An improvement of $5\sim 6$ orders of magnitude in the strain sensitivity is needed to reach $\Omega_{gw}h^2\approx 10^{-10}$, a possibility which may be achieved by further refinements to the prototype and/or construction of an array of such detectors \cite{Ballantini:2005am}. By way of comparison, we note that the best hope for observing a primordial gravitational wave background is currently provided by BBO, which has a lead time of at least 20-25 years. In this context hoping for a large extrapolation of detector technologies at high frequencies is perhaps not excessively optimistic.    In Figure~\ref{fig:sensitivityplot} we sketch the sensitivities of the leading interferometric detectors along with the expected stochastic background produced during preheating for the models we discuss in detail here.  

Finally, if one was to ever make a concerted attempt to detect a gravitational wave spectrum associated with preheating, one would need to be understand other potential sources that could supply a stochastic background of gravitational waves. This includes any first order phase transition in the early universe (such as the electroweak scale), or decays from cosmic strings. In addition, the presence of a rising component in the spectrum illustrates the dangers of using a (locally) positive spectral index of any detected stochastic gravitational wave background to rule out inflation in favour of alternative cosmogenesis ideas such as ekpyrosis \cite{Khoury:2001wf,Boyle:2003km} or pre-big bang scenarios \cite{Gasperini:1992em}.

 The potential for gravitational waves to provide a clean probe of inflation has rightfully drawn considerable attention, and strongly motivates attempts to detect the primordial gravitational spectrum. However, cosmic evolution is seldom tidy and gravitational waves are produced as long as large inhomogeneities are present. Preheating is a mechanism which will generate large inhomogeneities, and will necessarily be accompanied by the generation of a stochastic background of gravitational waves. The challenge now is to better determine their properties, and to assess possible strategies for their detection.

\section*{ Acknowledgments}

We  thank  Latham Boyle, Gary Felder, Gianluca Gemme, Tom Giblin,  Will Kinney,  Hiranya Peiris, Geraldine Servant, Igor Tkachev, and David Wands for a number of useful discussions. We are particularly indebted to Gary Felder and Igor Tkachev for their work on {\sc LatticeEasy\/}. This work is supported in part by  the United States Department of Energy, grant DE-FG02-92ER-40704.

\end{document}